\documentclass[letterpaper,11pt]{article}
\usepackage{amsmath, amsthm, amssymb}    
\usepackage{mathtools}
\usepackage{geometry}
\usepackage{graphicx}			
\usepackage{xcolor}				
\usepackage{tikz}
\usepackage{lipsum}
\usepackage[sort&compress,square,numbers]{natbib}
\usepackage[colorlinks=true, urlcolor=blue, citecolor=black, linkcolor=black]{hyperref}  
\usepackage{subcaption}			

\usepackage{titlesec}
\titlespacing\section{0pt}{10pt plus 4pt minus 2pt}{8pt plus 2pt minus 2pt}
\titlespacing\subsection{0pt}{10pt plus 4pt minus 2pt}{8pt plus 2pt minus 2pt}
\newcommand{\refEQ}[1]{Equation~(\ref{#1})}

\newcommand{\braces}[1]{\lbrace #1 \rbrace}
\newcommand{\brackets}[1]{\left( #1 \right)}
\newcommand{\reals}{\mathbb{R}}
\definecolor{cORANGE}{rgb}{0.784, 0.6, 0.}
\definecolor{cBLUE}{rgb}{0.275, 0.408, 0.635}
\definecolor{cORANGE2}{rgb}{1., 0.741, 0.254}

\newcommand{\sierpinski}{Sierpi\'{n}ski }

\NewDocumentCommand{\overarrow}{O{=} O{\uparrow} m}{%
	\overset{\makebox[0pt]{\begin{tabular}{@{}c@{}}#3\\[0pt]\ensuremath{#2}\end{tabular}}}{#1}
}
\NewDocumentCommand{\underarrow}{O{=} O{\downarrow} m}{%
	\underset{\makebox[0pt]{\begin{tabular}{@{}c@{}}\ensuremath{#2}\\[0pt]#3\end{tabular}}}{#1}
}

\title{Stitching Arrowhead Curves: Extending the \sierpinski Arrowhead Curve to Higher Dimensions}

\author{Eric Zimmermann\textsuperscript{1} and Stefan Bruckner\textsuperscript{2}
\vspace{10pt}\\
Institute for Visual \& Analytic Computing, University of Rostock, Germany \\ \textsuperscript{1}e.zimmermann@uni-rostock.de, \textsuperscript{2}stefan.bruckner@uni-rostock.de}
\date{}					

\begin{document}

\maketitle

\thispagestyle{empty}

\begin{abstract}
	The \sierpinski triangle and the \sierpinski arrowhead curve are both defined in dimension 2 and can be used to model the same fractal. While a natural extension of the triangular construction to arbitrary dimensions exists, an analogous extension of the curve representation does not. In this article, we analyze the properties of the two-dimensional \sierpinski arrowhead curve to formulate an extension to arbitrary dimensions based on reproduction rules. Building on this formulation, we demonstrate a way to visualize such curves in a comparative manner across levels. Finally, as geometric patterns have a long history in the arts, and especially in fashion, we exemplify this visualization approach in knitwear, specifically in the yoke of a sweater.
\end{abstract}

\section*{\sierpinski Triangle and Arrowhead Curves}

\begin{figure}[tbp]
	\centering
	\begin{minipage}[b]{0.15\textwidth} 
		\includegraphics[width=\textwidth]{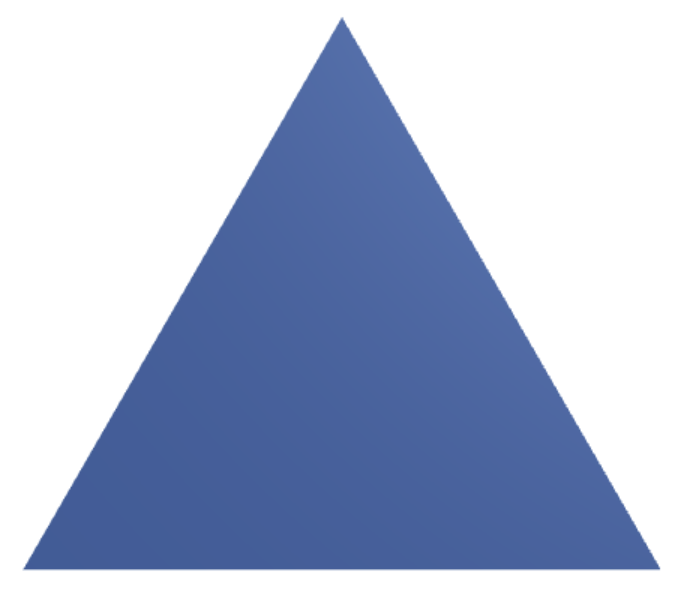}
		\includegraphics[width=\textwidth]{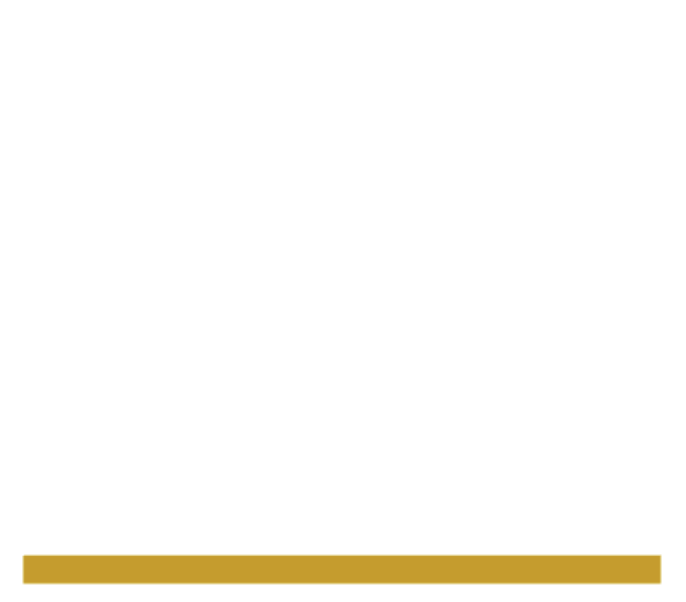}
		\subcaption{}
	\end{minipage}
	~
    \begin{minipage}[b]{0.15\textwidth} 
		\includegraphics[width=\textwidth]{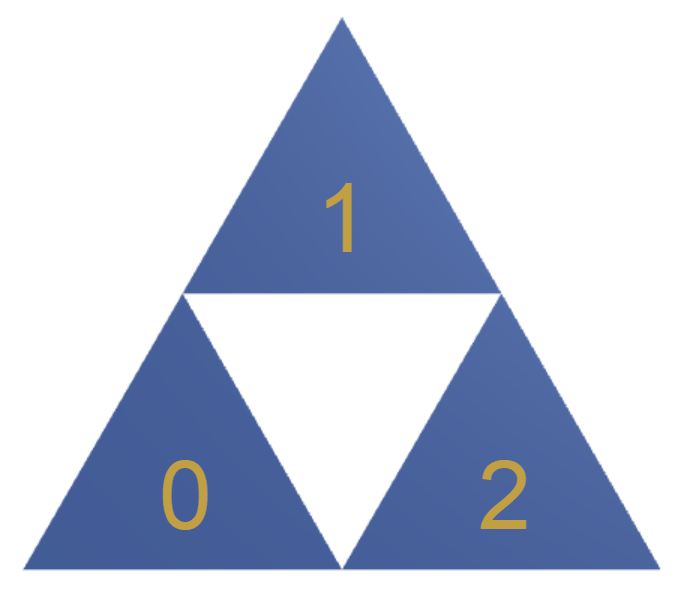}
		\includegraphics[width=\textwidth]{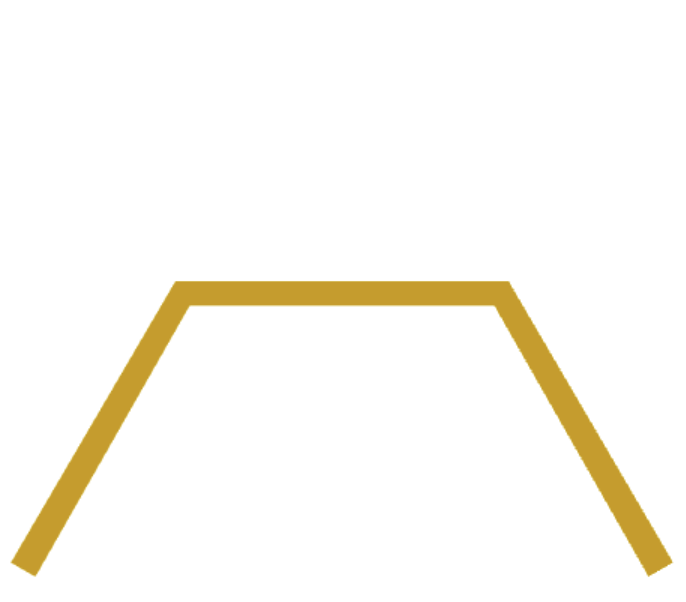}
		\subcaption{}
	\end{minipage}
	~
    \begin{minipage}[b]{0.30\textwidth}
		\includegraphics[width=\textwidth]{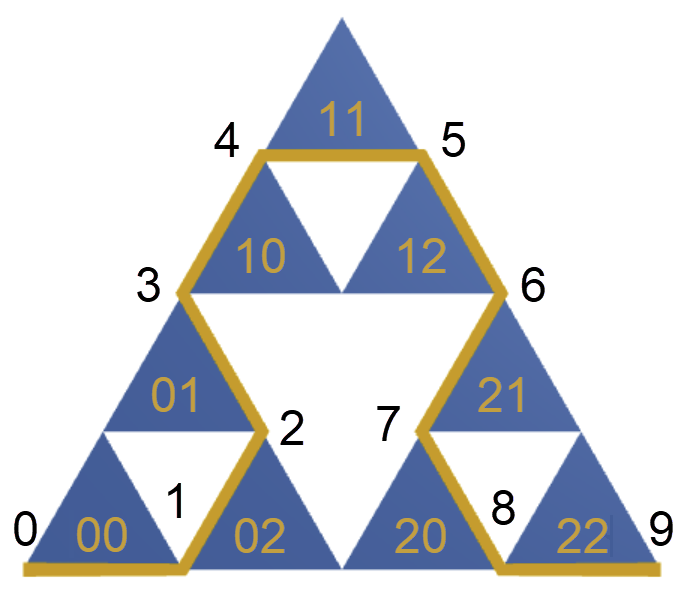}
		\subcaption{}
	\end{minipage}
	~
	\begin{minipage}[b]{0.15\textwidth}
		\includegraphics[width=\textwidth]{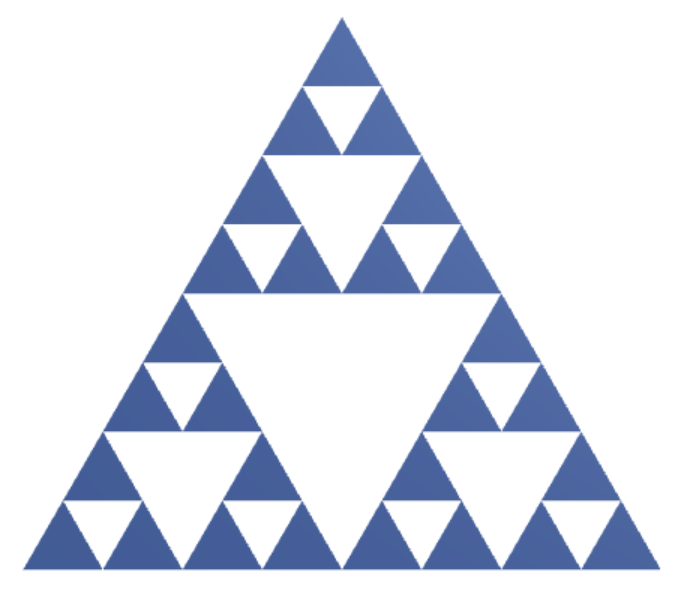}
		\includegraphics[width=\textwidth]{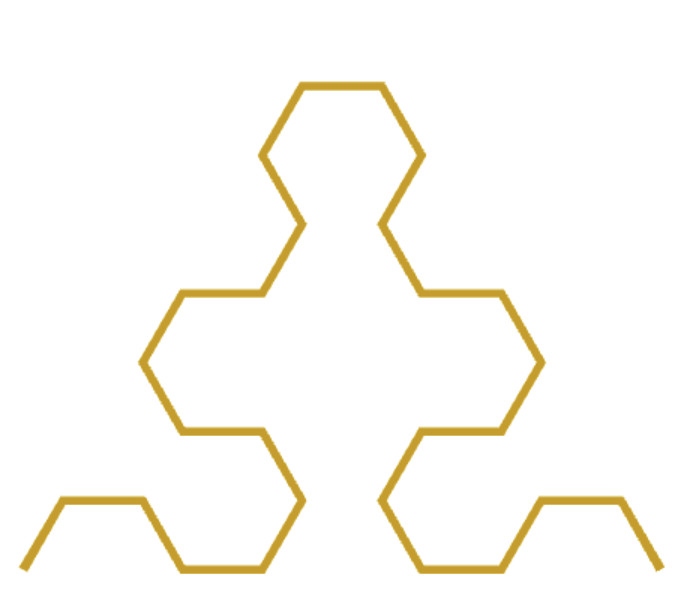}
		\subcaption{}
	\end{minipage}
	~
	\begin{minipage}[b]{0.15\textwidth}
		\includegraphics[width=\textwidth]{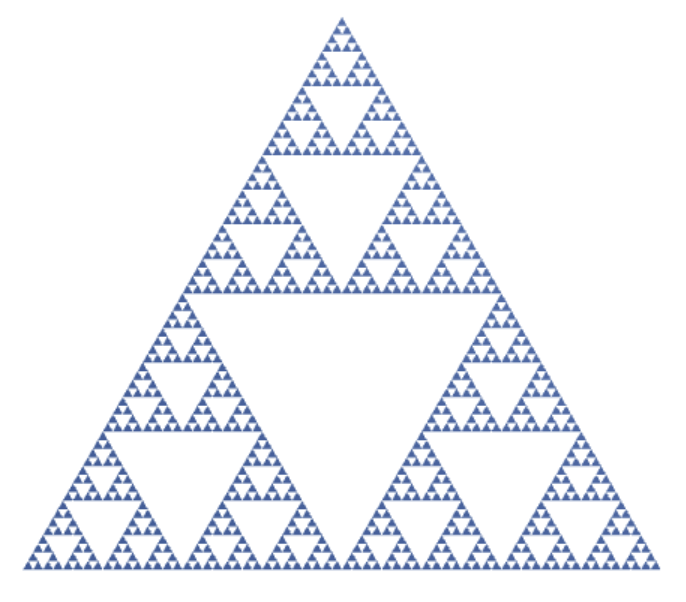}
		\includegraphics[width=\textwidth]{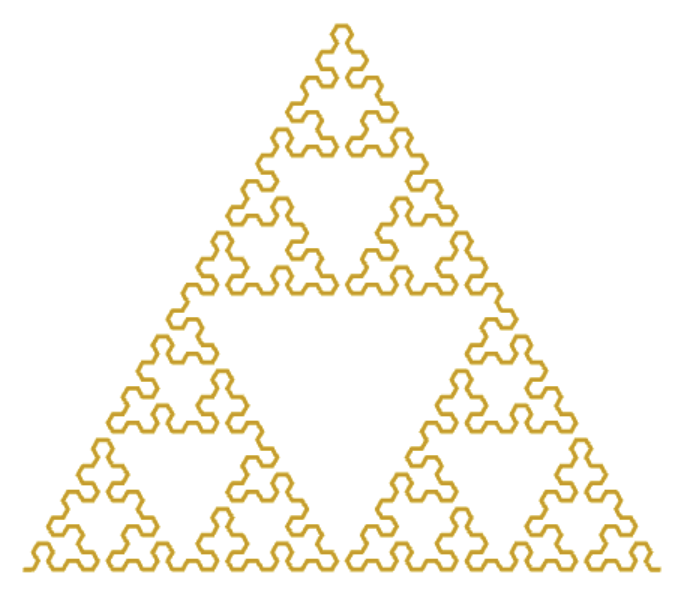}
		\subcaption{}
	\end{minipage}
	\caption{Levels 0, 1, 2, 3, and 6 of the \sierpinski triangle (top) and \sierpinski arrowhead curve (bottom) are shown. Labels for addresses (orange) are included in (b) and (c) and labels for curve points (black) are included in (c). Points of $\Delta^2$ are rotated before contraction for visual clarity such that $p_0$ and $p_2$ lie on the bottom.}
	\label{fig:introImages}
\end{figure}

The Sierpi\'{n}ski triangle is a well-known fractal defined within a triangle recursively subdivided into smaller triangles. One way to model it is to use an iterated function system (IFS). The following description is based on Taylor \cite{Taylor_Bridges25}, with generalizations and deviations noted where appropriate. The IFS for the Sierpi\'{n}ski triangle uses three contractions $f_0, f_1, f_2$ mapping from the plane to itself. We focus on the equilateral version, in which we start with a triangle $\Delta^2$ (already in generalized notation considered as a 2-simplex) as the convex hull of the three points $p_{2,0} = \brackets{1, 0}$, $p_{2,1} = \brackets{0, 1}$, and $p_{2,2} = \brackets{(1+\sqrt{3})/2, (1+\sqrt{3})/2}$
where $p_{d, i}$ represent the points for a given dimension $d \geq 2$ indexed by $i = 0, \ldots, d$. If the dimension is fixed and known we simply write $p_i$ which will apply for the remainder of this section, i.e., $d = 2$. The contraction maps use the triangle points as references, i.e.,
\[
    f_i\brackets{\brackets{x^1,x^2}} = \brackets{\frac{1}{2} \brackets{x^1 + p_{i}^1}, \frac{1}{2} \brackets{x^2 + p_{i}^2}} \quad \forall i = 0,1,2.
\]
Here superscripts refer to the respective coordinate of the point in question. For a given level $n \geq 1$, the triangle $\Delta^2$ is contracted via the composition $ f_{t_{n-1}} \circ f_{t_{n-2}}\circ \ldots \circ f_{t_0}(\Delta^2) $ with $t_i \in \braces{0,1,2}$, whereas each composition can be represented by an address $t_{n-1} t_{n-2}\ldots t_0$. When considering all permutations of the $n$ letters of such an address, we obtain all possible contractions for level $n$; the union of these contractions then represents the desired level $n$ of the \sierpinski triangle, cf. Figure \ref{fig:introImages} (c) for an example. Note that all illustrations are done using a publicly available web-application \cite{Demo}.

The \sierpinski arrowhead curve provides an alternative representation of the \sierpinski triangle \cite{Taylor_Bridges25}. It is a piecewise linear curve starting in $p_0$ going to $p_2$. In contrast to the contraction of triangles, the arrowhead curve consists of line segments that coincide with the edges of the contracted triangles at a given level, such that each triangle is visited exactly once, cf. Figure \ref{fig:introImages} (c). One way to describe this behavior
is by applying a \emph{reproduction rule} to the addresses when moving from one level to the next. This rule can be obtained from observing in which order the arrowhead curve visits the triangles and therefore the respective addresses. From Taylor \cite{Taylor_Bridges25} we learn that a block of three addresses reproduce into three blocks of three addresses as
\begin{equation}
	B a, B b, B c \to Baa, Bac, Bab, Bba, Bbb, Bbc, Bcb, Bca, Bcc,
	\label{eq:sec1:reproduction2D}
\end{equation}
where $a$, $b$, and $c$ denote the contraction indices 0, 1, and 2 in some order and $ B = t_{n-1} \ldots t_1 $ denotes an address from the former level. For example, letting $a=0$, $b=1$, and $c=2$, the transition from level 1 up to level 3 yields
\begin{align}
0,1,2 \quad \to \quad \underset{\star_1}{\underbrace{00, 02, 01 }}, \underset{\star_2}{\underbrace{10, 11, 12 }}, \underset{\star_3}{\underbrace{21, 20, 22}} \quad \to \quad
	\begin{split}
		  \,
        & 000, 001, 002, 020, 022, 021, 012, 010, 011, \\
        & 100, 102, 101 ,110, 111, 112, 121, 120, 122, \\
        & 211, 212, 210, 201, 200, 202, 220, 221, 222.
	\end{split}
	\label{eq:sec1:reproductionRule2DLevel1To3}
\end{align}
From level 1 to 2 the three contractions reproduce three blocks with three addresses each. Each block of three addresses in level 2 (marked with $\star$) reproduces three blocks each containing three new addresses. Thus, row $i$ in \refEQ{eq:sec1:reproductionRule2DLevel1To3} is the reproduction of $\star_i$ for $i=1,2,3$. Concerning preceding addresses, we have $B = \emptyset $ going from level 1 to 2, and $B = i$ for the respective reproduction $\star_i$ going from level 2 to 3 for $i=1,2,3$. An example from level 1 to 2 can be seen in Figure \ref{fig:introImages} (b) and (c). Note that addresses are read from right to left, reflecting the order in which the contractions are composed.

\section*{Extending Arrowhead Curves to Higher Dimensions}
The \sierpinski triangle and arrowhead curve are defined in dimension 2. Extending the triangular construction to higher dimensions is comparatively straightforward, and examples exist, such as the \sierpinski Tetrahedron in dimension 3 \cite{Hart_MathMuseum}.
For a given dimension $d \geq 2$, this construction requires $d+1$ points $p_{d,i} \in \reals^d$ forming a regular $d$-simplex $\Delta^d$, which can be defined as
\begin{equation}
	p_{d,i} = (0,\ldots, 0, \underset{i+1}{\underset{\uparrow}{1}}, 0, \ldots, 0) \quad \text{and} \quad p_{d,d} = \frac{1}{d} (1+\sqrt{d+1}) ( \underset{d \text{ times}}{\underbrace{1, \ldots, 1}} ) \quad \forall i = 0, \ldots, d-1.	
	\label{eq:sec2:simplexPoints}
\end{equation}
We will refer to these points as \emph{simplex points} in the following. Further, we need $d+1$ contraction maps
\begin{equation}
	f_i \brackets{ \brackets{x^1, \ldots, x^d}} = \frac{1}{2}\brackets{x^1 + p_{d, i}^1, \ldots, x^d + p_{d, i}^d} \quad \forall i = 0, \ldots, d,
	\label{eq:sec2:contractions}
\end{equation}
where superscripts refer to the respective coordinate of the point in question. As in the 2-dimensional case, we use an address $t_{n-1} t_{n-2}\ldots t_0$ with $t_j \in \braces{0, \ldots, d}$, to represent the composition $ f_{t_{n-1}} \circ f_{t_{n-2}}\circ \ldots \circ f_{t_0} $ contracting the $d$-simplex $\Delta^d$. Considering all permutations of the indices $t_j$ in an address of length $n$ yiels all required compositions. Their union represents the \emph{\sierpinski $d$-simplex of level $n$}. Examples for dimension~3 are shown in Figure \ref{fig:sec2:curves3Dexample}.

This naturally raises the question of how arrowhead curves can be defined in higher dimensions. Visual representations in dimension 3 already exist \cite{Dickau_Arrowhead3D}, but a systematic description that generalizes to arbitrary dimensions is still missing.

\subsection*{Reproduction Rules of Addresses}
We draw the following observations from the 2-dimensional case on how addresses reproduce from one level to the next and how the \sierpinski arrowhead curve behaves accordingly (cf. Equations~(\ref{eq:sec1:reproduction2D}), (\ref{eq:sec1:reproductionRule2DLevel1To3}), and Figure \ref{fig:introImages}). We focus on the transition from level 1 to 2, as the reproduction rule must already hold in this case.

\begin{itemize}
	\item[1.] We notice that level 1 is obtained without any reproduction, i.e., we simply apply the three contractions given on the left-hand side of the equation.
	\item[2.] At higher levels, we observe that each triangle (a \emph{super-triangle}) splits into three \emph{sub-triangles}, such that three addresses on the left reproduce nine addresses on the right.
	\item[3.] Since the curve starts in the lower left and terminates in the lower right, the first and last addresses on the right-hand side must be  $Baa$ and $Bcc$, respectively. Otherwise we are not able to contract $p_{2,0}$ and $p_{2,2}$ to the relevant positions.
	\item[4.] The curve visits each sub-triangle generated by a super-triangle before it continues to the next set of sub-triangles, e.g., it uses the order $10$, $11$, $12$, all of which came from the super-triangle with address $1$ from level 1. After that it follows with $21$, $20$, and $22$. Consequently, the order in which the curve traverses the super-triangles from level 1 is encoded by the addresses on the left, as well as by the second-to-last entry of each address on the right-hand side of the equation. Overall, the curve provides a sorting of triangles, cf. \sierpinski arrowhead order \cite{Taylor_Bridges25}.
	\item[5.] Closely related to the previous observation, whenever the curve transitions from one super-triangle to the next in level 1, the corresponding sub-triangle addresses share the same last and second-to-last entries, but in interchanged order.
\end{itemize}

We would like to extend this behavior to arbitrary dimensions while preserving the original $2$-dimensional case. In generalized form, we consider $d+1$ contraction maps with indices $a_j \in \braces{0, \ldots, d}$ and the following reproduction rule:
\begin{equation}
	\begin{split}
		Ba_0, \ldots, Ba_d \quad \to \quad 
		& Ba_0 a_0, \quad  \ldots,  \quad Ba_0 a_1, \\
		& Ba_1 a_0, \quad  \ldots,  \quad Ba_1 a_2, \\
		& \vdots  \\
		& Ba_d a_{d-1}, \quad \ldots, \quad Ba_d a_d
	\end{split}
	\overset{\text{simplified as}}{\longrightarrow}
	\begin{split}
		a_0, \ldots, a_d \quad \to \quad
		& a_0, \quad  \ldots,  \quad a_1, \\
		& a_0, \quad  \ldots,  \quad a_2, \\
		& \vdots  \\
		& a_{d-1}, \quad \ldots, \quad a_d.
	\end{split}
	\label{eq:sec2:reproductionRule}
\end{equation}
In relation to our observations above, we note the following: 1. Level 1 is obtained via the $d+1$ contractions $a_0, \ldots, a_d$; 2. Every $d+1$ addresses reproduce $(d+1)^2$ new ones; 3. The first and last address is fixed; 4.+5. Every $(d+1)$-th reproduced address and its successor are fixed.

The right-hand side in Equation~(\ref{eq:sec2:reproductionRule}) presents a simplified notation where we omit the preceding addresses $B$ as well as the second-to-last entries of the reproduced addresses. This simplification is possible because the preceding addresses are not required for understanding the reproduction process, although they must be carried along when generating objects in practise. The omitted second-to-last entries can be reconstructed from the sequence of the final entries on the left-hand side of either version. The other non-fixed reproduced addresses provide freedom of choice. The only restriction so far is that each index $a_0, \ldots, a_d$ needs to appear once as last entry in every address in each row on the right in Equation~(\ref{eq:sec2:reproductionRule}) to make sure that each contracted simplex gets visited by the curve. Typically and especially for later exploration, we only write down the right-hand side of the simplified version (Equation~(\ref{eq:sec2:reproductionRule})) to represent a reproduction rule, because it suffices to reconstruct and understand that it reproduces $a_0, \ldots, a_d$ from level 1 up to a specified level, as we will see later.

As the arrowhead curve shall visit each simplex and the reproduction rule is designed w.r.t. these curves, the reproduced addresses in a certain level provide all compositions of contractions necessary to end up with the \sierpinski $d$-simplex of level $n$. Hence, a reproduction rule provides a constructive mechanism for generating higher-dimensional \sierpinski simplices together with their associated arrowhead curves. Given a dimension $d \geq 2$, a level $n \geq 1$, and a reproduction rule according to Equation~(\ref{eq:sec2:reproductionRule}), we begin to sequentially reproduce the addresses starting with the initial block of $d+1$ addresses $a_0, \ldots, a_d$ ending up with $d+1$ blocks with $d+1$ addresses each. Then we continue to reproduce each of these blocks. With this procedure, the list of consecutive addresses (and thus contracted simplices) grows and we finish at the specified level, i.e., when all addresses have length $n$. We then apply each address of that list as a composition of contractions to the $3$-simplex using Equations~(\ref{eq:sec2:simplexPoints}) and (\ref{eq:sec2:contractions}).

Consider for example the objects illustrated in Figure~\ref{fig:sec2:curves3Dexample}. The following three levels of reproduction, each separated by an arrow indicate going from level 1 up to level 3:
\begin{eqnarray}
	0, 1, 2, 3 \quad \to \quad 
	\begin{split}
		&0,3,2,1, \\
		&0,1,3,2, \\
		&1,0,2,3, \\
		&2,1,0,3
	\end{split}
	\quad \to \quad 
	\begin{split}
		 & 0,1,2,3,0,3,1,2,3,0,2,1,2,3,0,1, \\
		 & 0,2,3,1,0,1,2,3,1,0,3,2,3,1,0,2,\\
		 & 1,3,2,0,1,0,3,2,0,1,2,3,2,0,1,3,\\
		 & 2,3,0,1,2,1,3,0,1,2,0,3,0,1,2,3.
	\end{split}
	\label{eq:sec2:reproductionExample3D}
\end{eqnarray}
They are given in simplified notation, i.e., only the last entry of each address is given. Actually the addresses in the center have length 2 and on the right have length 3. The reproduction rule is reflected in the first two columns (going from level 1 to 2). Afterwards each of the four rows with four addresses in the center reproduces one row on the right with 16 addresses each. Note that on level 3 we deviate from the row-wise representation used in Equation~(\ref{eq:sec2:reproductionRule}) to have a more compact form.

\subsection*{Arrowhead Curve Points}
\begin{figure}[tbp]
	\centering
	\begin{minipage}[b]{0.18\textwidth} 
		\includegraphics[width=\textwidth]{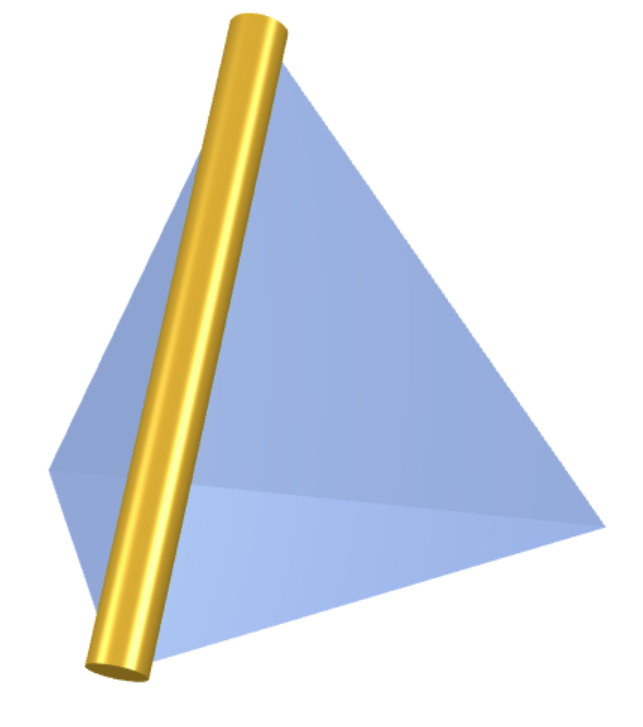}
	\end{minipage}
	~
	\begin{minipage}[b]{0.18\textwidth} 
		\includegraphics[width=\textwidth]{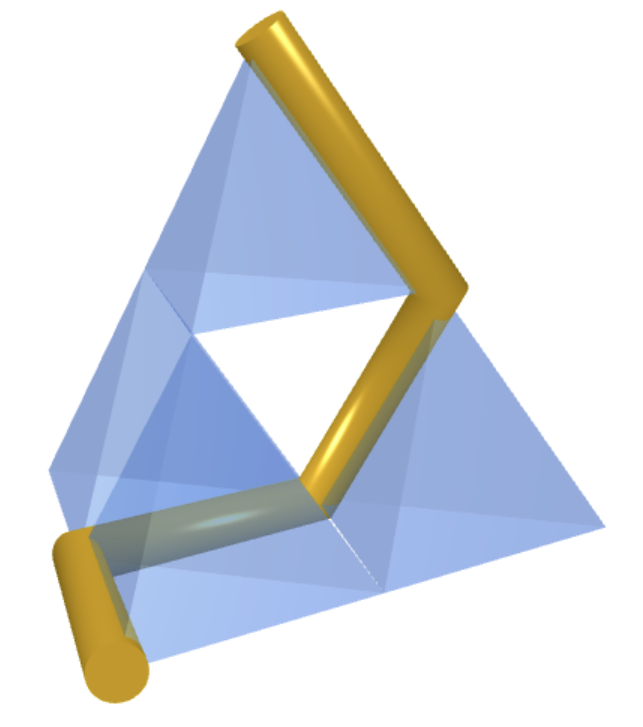}
	\end{minipage}
	~
	\begin{minipage}[b]{0.18\textwidth}
		\includegraphics[width=\textwidth]{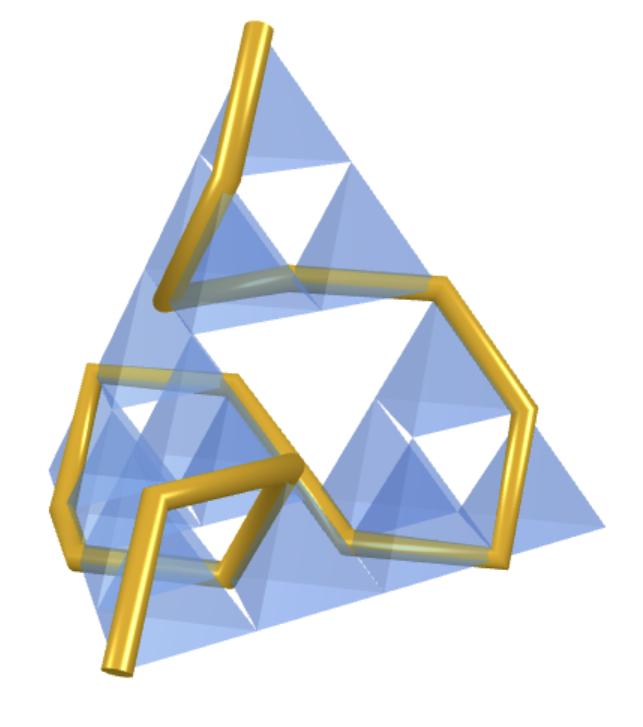}
	\end{minipage}
	~
	\begin{minipage}[b]{0.18\textwidth}
		\includegraphics[width=\textwidth]{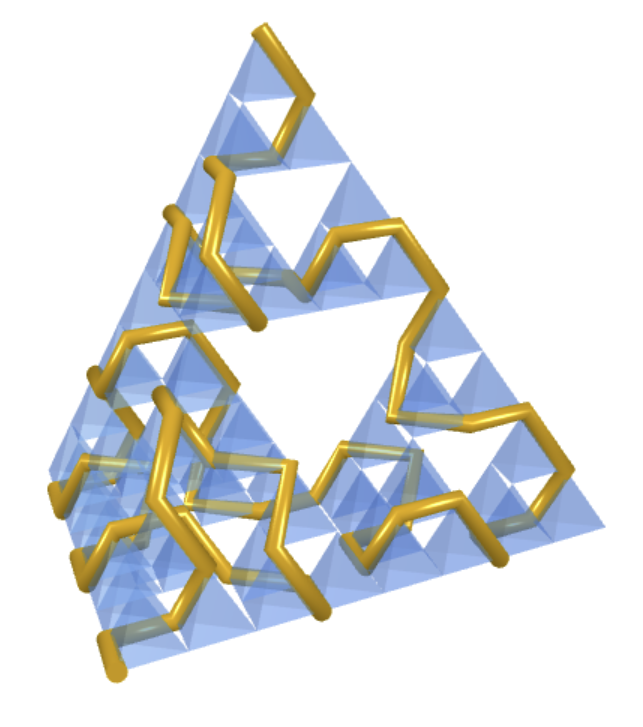}
	\end{minipage}
	~
	\begin{minipage}[b]{0.18\textwidth}
		\includegraphics[width=\textwidth]{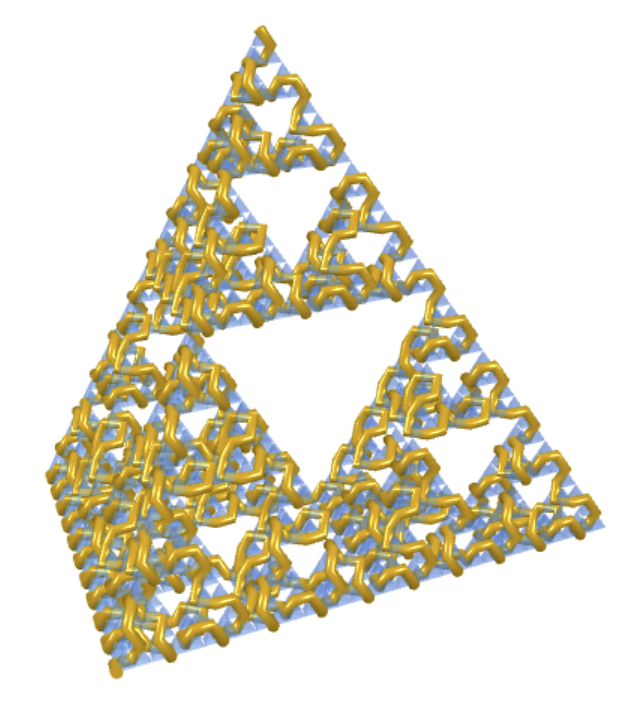}
	\end{minipage}
	\caption{Shows levels 0, 1, 2, 3, and 5 (f.l.t.r.) for the 3-arrowhead curve using Equations (\ref{eq:sec2:simplexPoints}), (\ref{eq:sec2:contractions}), and (\ref{eq:sec2:reproductionExample3D}) as well as the corresponding \sierpinski $3$-simplices (transparent).}
	\label{fig:sec2:curves3Dexample}
\end{figure}

So far, we have used the \sierpinski arrowhead order of triangles \cite{Taylor_Bridges25} to extend the sorting behavior to arbitrary dimensions,  controlled by a reproduction rule. This provides us the opportunity to generate the \sierpinski $d$-simplex of level $n$ w.r.t. a given rule, i.e., all contracted $d$-simplices, as each one is visited by the curve once. To obtain an explicit curve, however, we must identify the simplex edges that are incident to the curve. As before,  our goal is to reproduce the behavior of the original \sierpinski arrowhead curve in dimension~2. For this purpose, we simplify the notation and denote a contracted point by 
\[
t_{n-1} t_{n-2}\ldots t_0 (i)  \coloneq  f_{t_{n-1}} \circ f_{t_{n-2}}\circ \ldots \circ f_{t_0}(p_i) 
\]
in a fixed dimension $d$. We continue to use the terminology of super- and sub-triangles introduced in the discussion of reproduction rules. Analogous to the reproduction rules, we examine the curve behavior at levels~1 and 2 (cf. Figure \ref{fig:introImages} (b) and (c)), as these are the levels relevant in the reproduction. As the arrowhead curve is piecewise linear, we are interested in a sequence of points. At first, we observe that the initial and final points are $p_{0}$ and $p_{2}$, respectively. To account for the level we are currently in, we have to consider the contracted versions of the endpoints as $00(0)$ and $22(2)$, corresponding to points labeled 0 and 9 in Figure~\ref{fig:introImages}~(c), respectively. Following the curve across the triangles reveals two distinct behavioral patterns. In the first case, as long as the curve traverses all but the first sub-triangle generated by a super-triangle, the current address contracts the simplex point whose index equals the last entry of the previous address. For instance, the curve point with label 1 results from $02(0)$, while the point with label 2 results from $01(2)$. The second pattern occurs whenever the curve enters a sub-triangle belonging to a different super-triangle, i.e., when the curve transitions from one super-triangle to the next. In this case, the address contracts the simplex point whose index equals the last entry of the address itself. For example, the point with label 3 is obtained via $10(0)$. This process also applies to the very first sub-triangle at the start of the curve, as well as to the additional final point required to close the curve. The complete point sequences for the curves at levels 1 and 2  (cf. Figure~\ref{fig:introImages} (b) and (c)) are, in order, 
	\[
	0(0), 1(0), 2(1), 2(2) \quad \text{and} \quad 00(0), 02(0), 01(2), 10(0), 11(0), 12(1), 21(1), 20(1), 22(0), 22(2).
	\]

Now we extend these observations w.r.t. the reproduction rule given in Equation~(\ref{eq:sec2:reproductionRule}) to an arbitrary dimension. Recall that the reproduced addresses form a consecutive list reflecting the ordering of simplices induced by the curve. Consequently, each address has a position in that list. To generate a consecutive point list for the curve we traverse the list of addresses and distinguish two cases. First, if the address position modulo $d+1$ equals 1, the corresponding curve point is obtained by contracting the simplex point whose index equals the last entry of the address. When we think back to the row-wise representation in Equation~(\ref{eq:sec2:reproductionRule}), this is applied to each address being the first in each row. Second, for all remaining addresses, the curve point is obtained by contracting the simplex point whose index equals the last entry of the preceding address. When all addresses are processed, we need a final point for the curve which is obtained similar to the first case. So we use the very last address again and the simplex point with index equal to the last entry of this last address. The resulting list of points, defined by a reproduction rule for a given dimension $d$ and level $n$, forms a piecewise linear curve in $\reals^d$, which we call a \emph{$d$-Arrowhead curve of level $n$}.

Visual examples of curves in dimension 3 are illustrated in Figure \ref{fig:sec2:curves3Dexample} including levels 1 to 3 of the \sierpinski $3$-simplex given in Equation~(\ref{eq:sec2:reproductionExample3D}). The list of points, for instance, belonging to the $3$-Arrowhead curve of level $2$ are
\begin{align*}
	&00(0), 03(0), 02(3), 01(2), 10(0), 11(0), 13(1), 12(3), \\
	&21(1),20(1), 22(0), 23(2), 32(2), 31(2), 30(1), 33(0), \text{ and } 33(3).
\end{align*}

\section*{Visualization of $d$-Arrowhead Curves}
\begin{figure}[tbp]
	\centering
	\begin{minipage}[b]{0.2\textwidth} 
		\includegraphics[width=\textwidth]{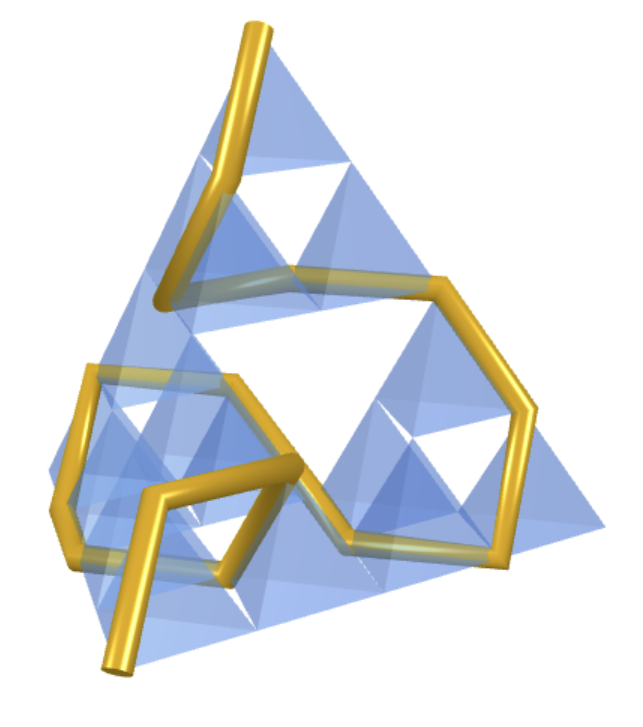}
	\end{minipage}
    ~
	\begin{minipage}[b]{0.2\textwidth}
		\includegraphics[width=\textwidth]{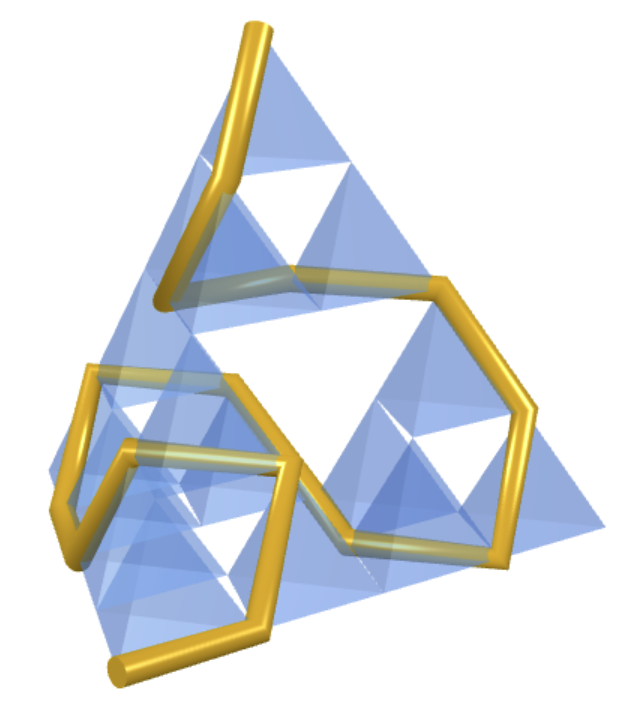}
	\end{minipage}
	~
	\begin{minipage}[b]{0.2\textwidth} 
		\includegraphics[width=\textwidth]{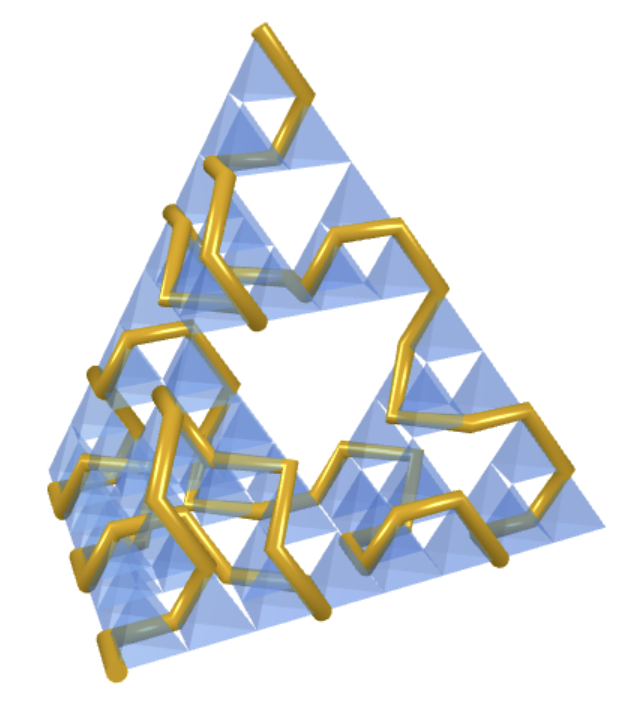}
	\end{minipage}
	~
	\begin{minipage}[b]{0.2\textwidth}
		\includegraphics[width=\textwidth]{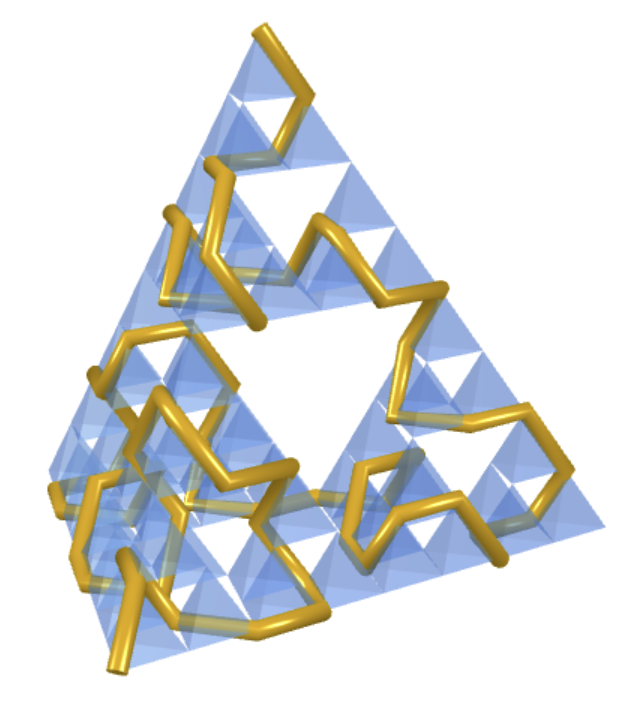}
	\end{minipage}
	\caption{Shows curves for rules $\mathfrak{R}_1$ and $\mathfrak{R}_2$ (f.l.t.r.) paired for levels 2 and 3.}
	\label{fig:sec3:differentRulesExamples3D}
\end{figure}

Now that we have extended the arrowhead curves to an arbitrary dimension w.r.t. a reproduction rule we might wonder how to make them understandable and not just a list of points in high-dimensional space. Visualizing such curves immediately raises challenges, as they must be projected to lower dimensions, typically 2 or 3. Identifying insightful properties and suitable projections of these curves is left to future work.
Beyond their intrinsic properties, these curves offer a variety of forms due to the freedom of choice in choosing reproduction rules for a fixed dimension $d$ and level $n$. This freedom arises from the flexible placements of indices in Equation~(\ref{eq:sec2:reproductionRule}). For a given dimension $d \geq 2$, only the first and last index in each row are fixed, while the remaining indices within each row can be freely permuted. With $d+1$ rows, we gain $(d-1)^{d+1}$ reproduction rules. As already noted, dimension 2 yields one rule, yet dimensions 3 and 4 provide the options for 16 and 243 rules, respectively. For example, two reproduction rules for $d=3$, one of which was already introduced in Equation~(\ref{eq:sec2:reproductionExample3D})), are
\begin{equation}
	\mathfrak{R}_1 = 0,3,2,1,0,1,3,2,1,0,2,3,2,1,0,3 \quad \text{and} \quad \mathfrak{R}_2 = 0,2,3,1,0,1,3,2,1,0,2,3,2,1,0,3,
	\label{eq:sec3:twoReproductionRuleExamples3D}
\end{equation}
both illustrated in Figure~\ref{fig:sec3:differentRulesExamples3D}. These examples reveal subtle variations in the spatial formations of the curves. While projections may still provide useful impressions in dimensions 3 and, to some extent, 4, such approaches become infeasible in higher dimensions. Moreover, increasing the refinement level further exacerbates this challenge, even in lower-dimensional projections.

\subsection*{Visualizing Binary Sequences}

\begin{figure}[tbp]
	\centering
	\includegraphics[width=\textwidth]{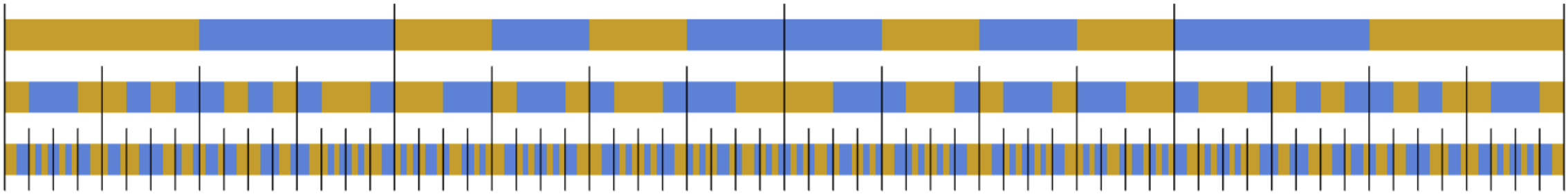}
	\includegraphics[width=\textwidth]{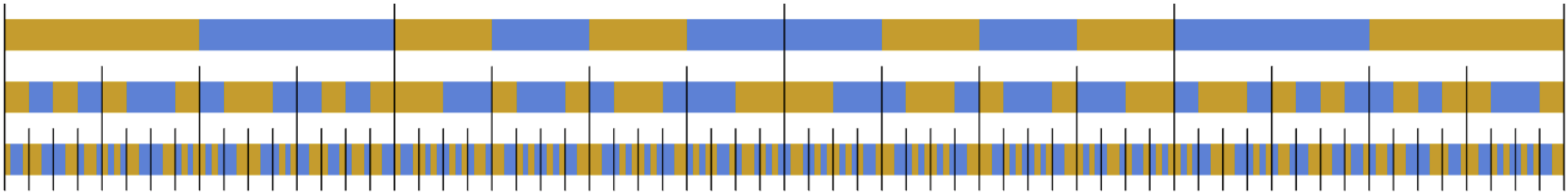}
	\caption{Visualizations of binary sequences for rules $\mathfrak{R}_1 $ (top) and $\mathfrak{R}_2 $, both for $d = 3$, each with levels 1 to 3 (top-down) with colored active (orange) and inactive (blue) indices and separators of $3$-simplices (black).}
	\label{fig:sec3:binarySequencesExamples3D}
\end{figure}

\begin{figure}[tbp]
	\centering
	\includegraphics[width=\textwidth]{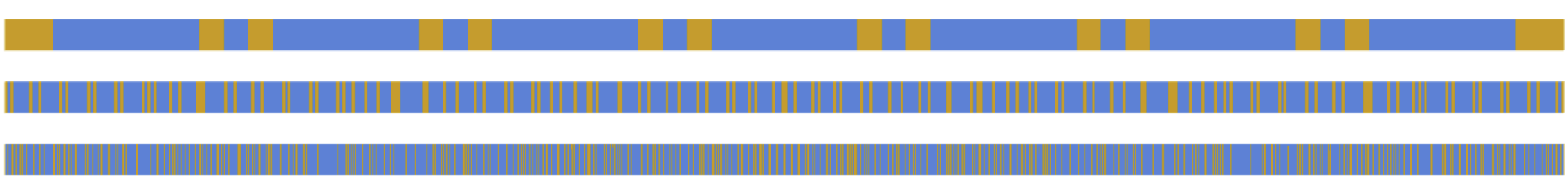}
	\caption{Visualization of binary sequences in dimension 7 with rule $\mathfrak{R}_3$ for levels 1 to 3 (top-down). An active index is colored in orange and an inactive in blue.}
	\label{fig:sec3:binarySequencesExample2}
\end{figure}

One option is to visualize $d$-arrowhead curves as binary sequences. An additional benefit is that multiple refinement levels can be arranged along a second dimension, allowing curves from different levels to be directly related.

Recall that the $d$-arrowhead curve induces an ordering of the $d$-simplices composing the \sierpinski $d$-simplex, and that the curve is incident to exactly one edge of each simplex, i.e., two of its points. Note further that each contracted $d$-simplex is a scaled copy of itself, consisting of contracted simplex points. Consequently, the ordering of simplex point indices is preserved between the original $d$-simplex and its contracted copies. Listing the simplex point indices and marking those incident to the curve as active, while marking the others as inactive, yields a binary sequence for the corresponding $d$-simplex. Concatenating these sequences, one per contracted $d$-simplex, and preserving the simplex order induced by the $d$-arrowhead curve yields a binary sequence representing the $d$-arrowhead curve at level $n$. Examples in dimension 3 and 7 can be seen in Figures~\ref{fig:sec3:binarySequencesExamples3D}~and~\ref{fig:sec3:binarySequencesExample2}, respectively. The latter example uses the reproduction rule
\begin{align*}
	\mathfrak{R}_3 \quad = \quad & 0, 2, 3, 4, 5, 6, 7, 1,
	0, 1, 3, 4, 5, 6, 7, 2,
	1, 0, 2, 4, 5, 6, 7, 3,
	2, 0, 1, 3, 5, 6, 7, 4, \\
	&3, 0, 1, 2, 4, 6, 7, 5, 
	4, 0, 1, 2, 3, 5, 7, 6, 
	5, 0, 1, 2, 3, 4, 6, 7, 
	6, 0, 1, 2, 3, 4, 5, 7.
\end{align*}
This way of representation allows us to compare and relate curves across levels and reproduction rules. For instance, in the 2-dimensional case we identify two aspects. The first is \emph{symmetry} w.r.t. the center, that means, the activity of the $i$-th position along a binary sequence equals the activity of the $(l-i+1)$-th position with $l$ denoting the length of the binary sequence starting with position 1. Note that symmetry is a property tied to the ordering of simplices by the curve and the type of visualization. However, not every reproduction rule yields a symmetric appearance. This can be seen in Figure~\ref{fig:sec3:binarySequencesExamples3D} for rule $\mathfrak{R}_2$ at level 2 in the sub-sequences representing the first and last $3$-simplices. Consequently, symmetry is a property tied to the reproduction rule and among the 16 possible rules in dimension 3, there are 4 which are symmetric. From here on we could generate all possible rules for a given dimension $d$ and determine those which cause symmetric binary sequences, which is left as a future direction.

The second aspect concerns \emph{self-similarity}. The \sierpinski triangle, and its $d$-dimensional analogue, are fractals and we can identify this property also for the representation via binary sequences. 
When we consider the result in Figure~\ref{fig:sec3:binarySequencesExamples3D}, for instance for rule $\mathfrak{R}_1$, we can see that the pattern (spanning four 3-simplices) in level~1 reoccurs in level 2 after applying isomorphisms on the simplex point indices. For instance, the first four elements in level 2 can be obtained mapping indices $0 \mapsto 0$, $1 \mapsto 3$, $2\mapsto2$, and $3\mapsto 1$. An interesting observation is that, given a reproduction rule, computationally identified isomorphisms are already present in the reproduction rule itself. Recalling the simplified version in Equation~(\ref{eq:sec2:reproductionRule}), 
we get an isomorphism per row on the right-hand side. For example we map the left-hand side component-wise to each element in the first row on the right-hand side. 

%
%
%
%
%
%
\section*{Binary Sequences in Knitting}

\begin{figure}[tbp]
	\centering
	\begin{minipage}[b]{0.3\textwidth} 
		\includegraphics[width=\textwidth]{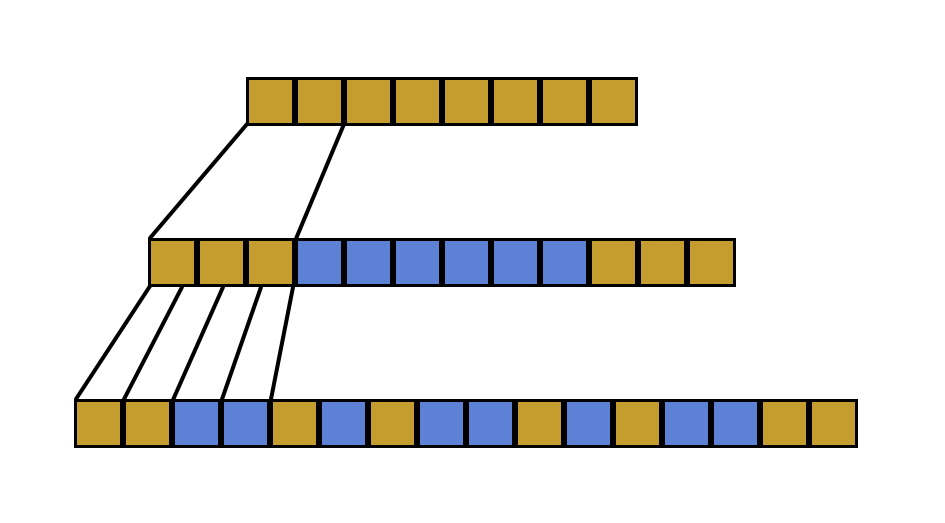}
		\subcaption{} 
		\label{fig:knitwear1}
	\end{minipage}
	~ 
	\begin{minipage}[b]{0.3\textwidth} 
		\includegraphics[width=\textwidth]{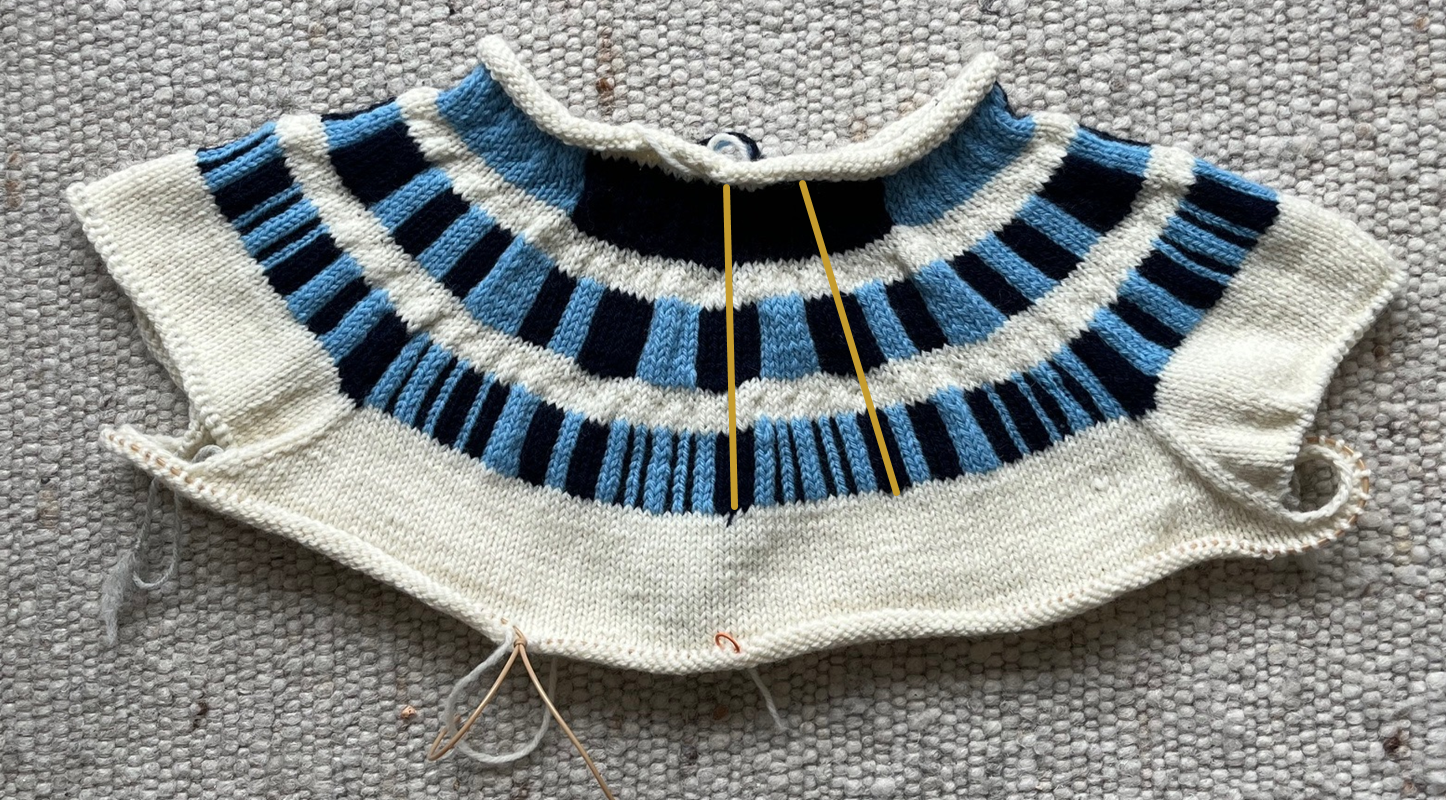}
		\subcaption{} 
		\label{fig:knitwear2}
	\end{minipage}
	~ 
	\begin{minipage}[b]{0.3\textwidth}
		\includegraphics[width=\textwidth]{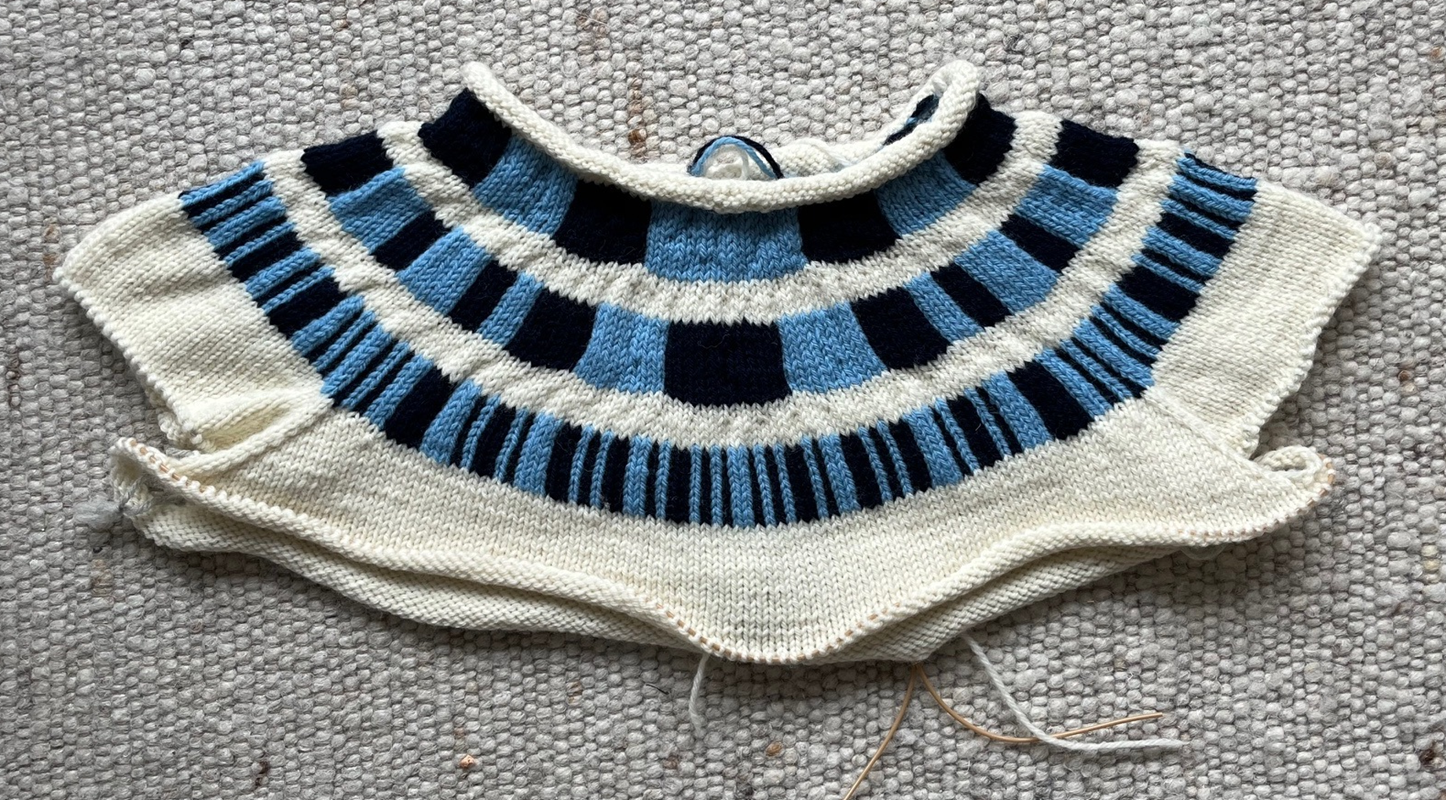}
		\subcaption{} 
		\label{fig:knitwear3}
	\end{minipage}
	\caption{Shows matching of stitches from level 1 to 3 for the rule $\mathfrak{R}_1$ in dimension $d = 3$ due to stitch increases (a) as well as back (b) and front (c) of the yoke. Highlight in (b) bounds stitch matching of the first sequence entry of level 1.}
	\label{fig:sec4:knitting}
\end{figure}

Geometric patterns have a long history in the arts, and are particularly prominent in fashion; a well-known example is the houndstooth pattern composed of alternating black and white checks. Monochromatic geometric patterns are likewise used to emphasize tangible structures, such as those found in the traditional Aran jumper originating from the Aran Islands off the west coast of Ireland. Although the underlying geometric elements in such patterns can be quite simple, symmetries enable aesthetically pleasing extensions and support scalability across different sizes. Knitting has even been proposed as a data visualization medium~\cite{Smit_Knitualization2021} that encodes information in textile form. This motivated us to use knitwear as a means to materialize the hierarchical and self-similar structure of the binary sequences derived from arrowhead curves.

We chose levels 1 to 3 of the symmetric binary sequence induced by rule $ \mathfrak{R}_1$ shown in Figure~\ref{fig:sec3:binarySequencesExamples3D} for visualization in knitwear, specifically in the yoke of a sweater. Yokes are often used to represent patterns and the stitch increases can be matched with the number of entries in the binary sequences (at least for one specific sweater size), i.e., at some point in the yoke we have 256 stitches matching the number of individual entries in the binary sequence of level 3 such that we have one stitch per entry, cf. Figure~\ref{fig:sec4:knitting} (b) and (c). To improve visibility and to clearly separate the three levels, we use nine rows per sequence and insert six rows in the main color (natural white) between successive levels. Active and inactive sequence entries are encoded using contrasting colors, dark and light blue, respectively. Since every colored sequence is surrounded by white, stitch increases are performed in every fourth row between the levels, resulting in a height of approximately 6.5\,cm per level, with the same number of white stitches carried above and below. The stich counts  per levels are 128 (level 1), 192 (level 2), and 256 (level 3). As these increases are distributed uniformly along each row, we matched the number of stitches used per entry in the binary sequence. For the first level, which contains 16 entries, this results in 8 stitches per entry. Figure~\ref{fig:sec4:knitting} (a) shows the first entry represented by 8 stitches in the first row, corresponding to the first point of the contracted $3$-simplex. Moving to level 2, each $3$-simplex reproduces four smaller $3$-simplices. In Figure~\ref{fig:sec4:knitting} (a), only representative stitch counts for the four corresponding points are shown in the second row. With 192 stitches and 64 sequence entries at level 2, this yields 3 stitches per entry. Accordingly, the single entry from level 1 splits into four entries, each now represented by three stitches. We apply the same procedure when transitioning from level 2 to level 3. Each entry from level 2 again splits into four entries, and the stitch increase results in four stitches per entry at level 3. In this way, the three stitches used at level 2 are consistently mapped to the four entries at level 3, each corresponding to one entry in the level-3 binary sequence. Following this procedure, we end up with a knitted yoke showing a $3$-arrowhead curve with levels 1 to 3, as shown in Figure~\ref{fig:sec4:knitting} (b) and (c).

%
%
%
%
%
%
\section*{Summary and Conclusions}
In this article we analyzed properties of the 2-dimensional \sierpinski arrowhead curve and its relation to the \sierpinski triangle to formulate an extension to arbitrary dimension. We visualized the resulting higher-dimensional curves as binary sequences and included them in knitwear as a representation of geometric patterns in fashion. An open question is whether the observations presented here imply additional constraints on admissible reproduction rules. Further insights may reduce the variety of admissible curves. Beyond the binary sequence representation, further research is needed to identify properties of high-dimensional curves and to develop suitable visualization strategies for them. Finally, creating additional artifacts that incorporate such patterns could further support their use in artistic practice and make these abstract structures tangible for educational purposes.

\section*{Acknowledgements}
We are grateful for the idea, discussions, and knitting of the yoke done by Eric's wife Anne Zimmermann.

    
{
\setlength{\baselineskip}{12pt} 
\raggedright				
\bibliographystyle{plain}
\bibliography{article}
} 
   
\end{document}